\documentclass[twocolumn,showpacs,preprintnumbers,amsmath,amssymb,prb,superscriptaddress]{revtex4}
\usepackage{graphicx}

\begin{document}

\title{\bf Continuous Time Quantum Monte Carlo method for fermions}

\author{A. N. Rubtsov}
\affiliation{Department of Physics, Moscow State University,
119992 Moscow, Russia}
\author{V. V. Savkin}
\affiliation{Institute of Theoretical
Physics, University of Nijmegen, 6525 ED Nijmegen, The
Netherlands}
\author{A. I. Lichtenstein}
\email{aichten@physnet.uni-hamburg.de}
\affiliation{Institute of Theoretical Physics, University of
Hamburg, 20355 Hamburg, Germany} \pacs{71.10.Fd, 71.27.+a,
02.70.Ss}

\begin{abstract}
We present a numerically exact continuous-time Quantum Monte Carlo
algorithm for fermions with a general interaction non-local in
space-time. The new determinantal grand-canonical scheme is based
on a stochastic series expansion for the partition function in the
interaction representation. The method is particularly applicable
for multi-band, time-dependent correlations since it does not
invoke the Hubbard-Stratonovich transformation. The test
calculations for exactly solvable models, as well results for the
Green function and for the time-dependent susceptibility of the
multi-band super-symmetric model with a spin-flip interaction are
discussed.
\end{abstract}

\maketitle

\section{Introduction}

The variety of quantum Monte Carlo (QMC) methods is the most
universal tool for the numerical study of quantum many-body
systems with strong correlations. So-called determinental Quantum
Monte Carlo (QMC) scheme for fermionic systems appeared more than
20 years ago \cite{Scalapino, Blankenbecler, Hirsch, HirschPH}.
This scheme has became standard for the numerical investigation of
physical models with strong interactions, as well as for quantum
chemistry and nanoelectronics. Although the first numerical
attempts were made for model Hamiltonians with local interaction,
the real systems are described by the many-particle action of a
general form. For example many non-local matrix elements of the
Coulomb interaction do not vanish in the problems of quantum
chemistry \cite{qchem} and solid state physics \cite{qmcpw}. For
realistic description of Kondo impurities like a cobalt atom on a
metallic surface it is of crucial importance to use the spin and
orbital rotationally invariant Coulomb vertex in the
non-perturbative investigation of electronic structure. The
recently developed dynamical mean-field theory (DMFT) \cite{DMFT}
for correlated materials introduces a non-trivial
frequency-dependent bath Green function. The extension \cite{TDU}
of the theory deals with an interaction that is non-local in time.
Moreover, the same frequency dependent single-electron
Green-function and retarded electron-electron interaction
naturally appear in any electronic subsystem where the rest of
system is integrated out. An interesting non-local effect due to
off-diagonal exchange interactions may be responsible for the
correlated superconductivity in the doped fullerens
\cite{Fullerens}. It is worth noting that the exchange mechanism
often has an indirect origin (like the super-exchange) and the
exchange terms can therefore be retarded.

The  determinantal grand-canonical auxiliary-field scheme
\cite{Scalapino, Blankenbecler, Hirsch, HirschPH} is extensively
used for interacting fermions, since other known QMC schemes (like
stochastic series expansion in powers of Hamiltonian \cite{SSE} or
worm algorithms \cite{Worm}) suffer from an unacceptably bad sign
problem for this case. The following two points are essential for
the determinantal QMC approach: first, the imaginary time is
artificially discretized, and the Hubbard-Stratonovich
transformation \cite{Hubbard} is performed to decouple the
fermionic degrees of freedom. After the decoupling, fermions can
be integrated out, and Monte Carlo sampling should be performed in
the space of auxiliary Hubbard-Stratonovich fields. Hirsch and Fye
\cite{Hirsch} proposed a so-called discrete Hubbard-Stratonovich
transformation to improve the efficiency of original scheme. It is
worth noting that for a system of $N$ atoms the number of
auxiliary fields scales $\propto N$ for the local (short-range)
interaction and as $N^2$ for the long-range one. This makes the
calculation rather ineffective for the non-local case. In fact the
scheme is developed for the local interaction only.

The problem of systematic error due to the time discretization was
addressed in several works. For bosonic quantum systems, the
continuous time loop algorithm \cite{Beard}, worm diagrammatic
world line Monte Carlo scheme \cite{Worm}, and continuous time
path-integral QMC\cite{Kornilovitch} overcame this problem.
Recently a continuous-time modification of the fermionic QMC
algorithm was proposed \cite{CT}. It is based on a series
expansion for the partition function in the powers of interaction.
The scheme is free from time-discretization errors but the
Hubbard-Stratonovich transformation is still invoked. Therefore
the number of auxiliary fields scales similarly as the discrete
scheme, so that the method remains local.

The most serious problem of the QMC simulation for large systems
and small temperatures is the sign problem \cite{SignNP} resulting
in the exponential growth of computational time. This is a
principal drawback of the QMC scheme \cite{SignNP}, but it is
system dependent. For relatively small clusters, in particular for
the local DMFT scheme, the sign problem is not crucial
\cite{DMFT,clusterDMFT}. If we consider any subsystem obtained by
integrating out the rest of the system, the Gaussian part as well
as the interaction for the new effective action are non-local in
space-time. Unfortunately, as we pointed out, the non-locality of
the interaction hampers the calculation because it is hard to
simulate systems with a large number of auxiliary spins. It is
nearly impossible to simulate a system with interactions that are
non-local also in time, when the number of spins is proportional
to $(\beta N/\delta \tau)^2$ ($\beta$ is inverse temperature, and
$\delta \tau$ is a time-slice).

Recent developments in the field of interacting fermion systems
\cite{newDMFT} clearly require the construction of a new type of
QMC scheme suitable for non-local, time-dependent interactions. In
this paper we present a continuous-time quantum Monte Carlo
(CT-QMC) algorithm which  does not introduce any auxiliary-field
variables. The principal advantages of the present algorithm are
related to the different scaling of the computational time for
non-local interactions. The scheme is particularly suitable for
general multi-orbital Coulomb interactions. The paper is aimed at
a general description of the algorithm and the estimation of the
computation complexity. We present the results for test systems to
show an adequacy of the method. Moreover, an analysis of a
non-trivial multi-band rotationally-invariant model with a
time-dependent interaction is performed. This model demonstrates
the main advantages of the numerical scheme. The paper is
organized as follows. In Section II we discuss a general
formalism. Section III contains several applications of CT-QMC
scheme for simple systems in comparison with the exact solutions
and results of the super-symmetric multi-band impurity problem.
The conclusions are given in the Section IV.

\section{Formalism}

\subsection{General principles}

One can consider the partition function for the system with a pair
interaction in the most general case which has the following form:
\begin{widetext}
\begin{eqnarray}\label{S}
&Z= \mathrm{Tr} T e^{-S},\\
&S=\int \int   t_r^{r'}  c^\dag_{r'} c^{r} dr  dr' + \int \int
\int \int w_{r_1 r_2} ^{r_1' r_2'} c^\dag_{r_1'} c^{r_1}
c^\dag_{r_2'} c^{r_2} dr_1 dr_1' dr_2 dr_2'. \nonumber
\end{eqnarray}
\end{widetext}
Here $T$ is a time-ordering operator, $r=\{\tau,s,i\}$ is a
combination of the continuous imaginary-time variable $\tau$, spin
orientation $s$ and the discrete index $i$ numbering the
single-particle states in a lattice. Integration over $d r$
implies the integral over $\tau$ and the sum over all lattice
states and spin projections: $\int dr \equiv \sum_i \sum_s
\int_0^\beta d\tau$.

One can now split $S$ into two parts: the unperturbed action $S_0$
in a Gaussian form and an interaction $W$. We introduce as well an
additional quantity $\alpha_{r'}^{r}$, which can be in principle a
function of time, spin, and the number of lattice state. The
functions $\alpha_{r'}^{r}$ will later help us to minimize the
sign problem and to optimize the algorithm. Thus up to an additive
constant we have:
\begin{widetext}
\begin{eqnarray}\label{S0}
&S=S_0+W,\\
&S_0=\int \int   \left( t_r^{r'}+\int \int \alpha^{r_2}_{r_2'}
(w_{r r_2} ^{r' r_2'} + w_{r_2 r} ^{r_2' r'}) dr_2 dr_2' \right) ~
c^\dag_{r'} c^{r} ~ dr  dr', \nonumber \\
&W=\int \int \int \int
w_{r_1 r_2} ^{r_1' r_2'} (c^\dag_{r_1'}
c^{r_1}-\alpha^{r_1}_{r_1'}) (c^\dag_{r_2'}
c^{r_2}-\alpha^{r_2}_{r_2'}) dr_1 dr_1' dr_2 dr_2'. \nonumber
\end{eqnarray}

Now we switch to the interaction representation and make the
perturbation series expansion for the partition function $Z$
assuming $S_0$ as an unperturbed action:

\begin{eqnarray}\label{Z}
&Z=\sum\limits_{k=0}^\infty Z_k=\sum\limits_{k=0}^\infty \int dr_1
\int d r_1' ... \int dr_{2k} \int dr_{2k}' \Omega_k (r_1, r_1',
..., r_{2k}, r_{2k}'),
\\ \nonumber &\Omega_k=Z_0 \frac{(-1)^k}{k!}w_{r_1 r_2}^{r_1' r_2'} \cdot ...
\cdot w_{r_{2k-1} r_{2k}}^{r_{2k-1}' r_{2k}'} D^{r_1 r_2 ...
r_{2k}}_{r_1' r_2' ... r_{2k}'}.
\end{eqnarray}
\end{widetext}
Here $Z_0= \mathrm{Tr} T e^{-S_0}$ is a partition function for the
unperturbed system and
\begin{equation}\label{D}
D^{r_1 ... r_{2k}}_{r_1' ... r_{2k}'} = <T (c^\dag_{r_1'}
c^{r_1}-\alpha^{r_1}_{r_1'}) \cdot ... \cdot (c^\dag_{r_{2k}'}
c^{r_{2k}}-\alpha^{r_{2k}}_{r_{2k}'})>.
\end{equation}
Hereafter the triangle brackets denote the average over the
unperturbed system (for arbitrary operator $A$: $<A> = Z_0^{-1}
\mathrm{Tr} T A e^{-S_0}$). Since $S_0$ is Gaussian, one can apply
Wick's theorem to transform (\ref{D}). Thus $D^{r_1 ...
r_{2k}}_{r_1' ... r_{2k}'}$ is a determinant of a $2k \times 2k$
matrix which consists of the two-point bare Green functions $g_{0
r'}^r=<T c^\dag_{r'} c^{r}>$ at $\alpha_{r'}^{r}=0$. Obviously,
for non-zero $\alpha_{r'}^{r}$
\begin{equation}\label{det}
D^{r_1 r_2 ... r_{2k}}_{r_1' r_2' ... r_{2k}'}= \det||
g^{r_i}_{0r_j'}- \alpha^{r_i}_{r_j'} \delta_{ij} ||; i,j=1,...,2k,
\end{equation}
where $\delta_{ij}$ is a delta-symbol.

Now we can express the two-point Green function for the system
(\ref{S}) using the perturbation series expansion (\ref{Z}). It
reads:
\begin{widetext}
\begin{equation}\label{G}
G_{r'}^r \equiv Z^{-1} <T c^\dag_{r'} c^{r} e^{-W}>=\sum_k \int
dr_1 \int d r_1' ... \int dr_{2k} g_{r'}^r (r_1, r_1', ...,
r_{2k}') \Omega_k(r_1, r_1', ..., r_{2k}'),
\end{equation}
\end{widetext}
where $g_{r'}^r(r_1, r_1', ..., r_{2k}')$ denotes the Green
function for a general term of the series
\begin{eqnarray}\label{g}
&g_{r'}^r(r_1, r_1', ..., r_{2k}')=(D^{r_1 ... r_{2k}}_{r_1' ...
r_{2k}'})^{-1} \times \\
&\times<T c^\dag_{r'} c^{r} (c^\dag_{r_1'}
c^{r_1}-\alpha_{r'_1}^{r_1})\cdot...\cdot(c^\dag_{r_{2k}'}
c^{r_{2k}}-\alpha_{r'_{2k}}^{r_{2k}})>. \nonumber
\end{eqnarray}
Similarly, one can write formulas for other averages, for example
the two-particle Green function.

An important property of the above formulas is that the integrands
stay unchanged under the permutations $r_{i}, r_{i'}, r_{i+1},
r_{i'+1} \leftrightarrow r_{j}, r_{j'}, r_{j+1}, r_{j'+1}$ with
any $i,j$. Therefore it is possible to introduce a quantity $K$,
which we call "state of the system" and is a combination of the
perturbation order $k$ and an {\it unnumbered set} of $k$ tetrades
of coordinates. Now, denote $\Omega_K=k! \Omega_k$, where the
factor $k!$ reflects all possible permutations of the arguments.
For the Green functions, $k!$ in the nominator and denominator
cancel each other, so that $g_K=g_k$.

In this notation,
\begin{eqnarray}\label{Klanguage}
    &Z=\int \Omega_K D [K], \\   \nonumber
    &G_{r'}^r = Z^{-1} \int g_K \Omega_K D[K],
\end{eqnarray}
where $\int D[K]$ means the summation by $k$ and integration over
all possible realizations of the above-mentioned unnumbered set at
each $k$. One can check that the factorial factors are indeed
taken into account correctly with this definition.

\subsection{Convergence of the perturbation series}

It is important to notice that the series expansion for an
exponent {\it always} converges for the finite fermionic systems.
A mathematically rigorous proof can be constructed for Hamiltonian
systems. Indeed, the many-body fermionic Hamiltonians $H_0$ and
$W$ have a finite number of eigenstates that is $2^N$, where $N$
is the total number of electronic spin-orbitals in the system. Now
one can observe that $\Omega_k<const \cdot W_{max}^k$, where
$W_{max}$ is the eigenvalue of $W$ with a maximal modulus. This
proves convergence of (\ref{Z}), because the $k!$ in the
denominator grows faster than the numerator. In our calculations
for the non-Hamiltonian systems we also did not observe any
indications of the divergence.

The crucial point of the proof is the finiteness of the number of
states in the system. This is a particular peculiarity of
fermions. For bosons, on other hand, one deals with a Hilbert
space of an infinite dimensionality. Therefore series like
(\ref{Z}) are known to be divergent even for the simplest case of
a single classical anharmonic oscillator \cite{IZ}. It is
important to keep this in mind for possible extensions of the
algorithm to the electron-phonon system and to the field models,
since these systems are characterized by an infinite-order phase
space.

It is also important to note that this convergence is related to
the choice of the type of series expansion. Indeed, the series
(\ref{Z}) contains {\it all} diagrams, including disconnected. In
the analytical diagram-series expansion disconnected diagrams drop
out of the calculation and the convergence radius for
diagram-series expansion differs from that of Eq.(\ref{Z}).

For the purpose of real calculation, it is desirable to estimate
which values of $k$ contribute the most to $Z$. It follows from
the formula (\ref{Z}) that
\begin{equation}\label{Kaver}
    <k>=<W>.
\end{equation}
This formula gives also a simple practical recipe for how to
calculate $<W>$. For example, in an important case of the on-site
Coulomb interaction, it gives information about the local
density-density correlator.

\subsection{Random walk in $K$-space}

Although formula (\ref{Klanguage}) looks rather formal, it exactly
corresponds to the idea of the proposed CT-QMC scheme. We simulate
a Markov random walk in a space of all possible $K$ with a
probability density $P_K \propto |\Omega_K|$ to visit each state.
If such a simulation is implemented, obviously
\begin{equation}\label{Avs}
    G_{r'}^r=\overline{s g_{r'}^r}/\overline{s}
\end{equation}
The overline here denotes a Monte Carlo averaging over the random
walk, and $\overline{s}=\overline {\Omega_K/|\Omega_K|}$ is an
average sign.

Two kinds of trial steps are necessary: one should try either to
increase or to decrease $k$ by 1, and, respectively, to add or to
remove the corresponding tetrad of "coordinates".

Suppose that we perform incremental and decremental steps with an
equal probability. Consider a detailed balance between the states
$K$ and $K'$, where $K'$ is obtained by an addition of certain
tetrad $r_{2k+1}, r'_{2k+1}, r_{2k+2}, r'_{2k+2}$ to $K$. It
should be noted that  $P(K)$ and $P(K')$ appear under integrals of
different dimensionality, respectively $k$ and $k+4$. Therefore it
is more correct to discuss the detailed balance between the state
$K$ and all $K'$ with $r_{2k+1}, r'_{2k+1}, r_{2k+2}, r'_{2k+2}$
corresponding to a certain domain $d^4 r$. The detailed balance
condition reads
\begin{equation}\label{Detail}
    \frac{P_{K\to K'}}{P_{K'\to K}}=\frac{P_{K'}~ d^4r}{P_K},
\end{equation}
where $P_{K\to K'}$ is a probability to arrive in $K'$ after a
single MC step from $K$.

In the incremental steps the proposition for the four new points
should be generated randomly. Denote the probability density in
this generation $p(r_{2k+1}, r'_{2k+1}, r_{2k+2}, r'_{2k+2})$. If
this step is accepted with a conditional probability $p_{K\to
K'}$, then
\begin{equation}
P_{K\to K'}=p_{K\to K'} p(r_{2k+1}, r'_{2k+1}, r_{2k+2},
r'_{2k+2}) d^4 r.
\end{equation}

For the decremental steps, it is natural to pick randomly one of
the existing tetrades and consider its removal. So,
\begin{equation}
P_{K'\to K}=p_{K'\to K}/(k+1).
\end{equation}

Therefore, one obtains the condition for acceptance probabilities:
\begin{equation}\label{accept}
    \frac{p_{K'\to K}}{p_{K\to K'}}=\left|\frac{\Omega_K}
    {\Omega_{K'}} \right| (k+1) p(r_{2k+1}, r'_{2k+1}, r_{2k+2}, r'_{2k+2}).
\end{equation}

In principle, one can choose different $p(w_{r_{2k+1}
r_{2k+2}}^{r_{2k+1}' r_{2k+2}'})$, it is important only to
preserve (\ref{accept}). We propose to use
\begin{eqnarray}
&p=
||w||^{-1} |w_{r_{2k+1} r_{2k+2}}^{r_{2k+1}' r_{2k+2}'}| \\
\nonumber &||w||={\int \int \int \int  |w_{r R}^{r' R'}|} dr dR
dr' dR'
\end{eqnarray}
to generate new points in the incremental steps. Then the standard
Metropolis acceptance criterion can be constructed using the ratio
\begin{equation}\label{prob+}
    \frac{||w||}{k+1} \cdot
    \left| \frac{D^{r_1  ... r_{2k+2}}_{r_1' ... r_{2k+2}'} }
    {D^{r_1  ... r_{2k}}_{r_1' ... r_{2k}'} } \right|.
\end{equation}
for the incremental steps and its inverse for the decremental
ones.

In general, one may want also to add-remove several tetrades
simultaneously. A thus organized random walk is illustrated by
Figure 1. The same Figure presents a typical distribution diagram
for a perturbation order $k$ in QMC calculation.

\begin{figure}
\includegraphics[width=\columnwidth]{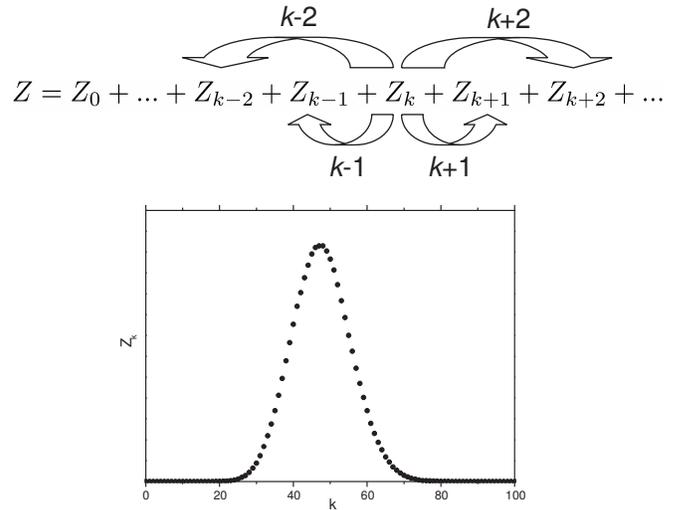}
\caption{\label{walk} Schematic picture of random walks in the
space of $k; r_1, r_1', ..., r_{2k}, r_{2k}'$ according to
perturbation series expansion (\ref{Z}) and an example of the
histogram for the perturbation order $k$.}
\end{figure}

\subsection{A fast-update of Green function matrix}

The most time consuming operation of the algorithm is the
calculation of the ratio of determinants and Green-function
matrix. It's necessary for calculation of MC weights as well as
for Green function.

There exist so called fast-update formulas for calculation of the
ratio of determinants and Green-function matrix. Usual procedure
takes $N^3$ operations, while the fast-update technique allows one
to perform $N^2$ or less operations, where $N$ is a matrix size.

Our derivation of the fast-update formulas is a generalization of
the Shermann-Morrison scheme for the determinatal QMC. Usually,
the two types of steps ($k\to k+1$ and $k\to k-1$) are sufficient.
However, the steps $k \to k \pm 2$ can be also employed in certain
cases (see later examples), so we present here also formulas for
that case.

We use the following notation to derive the fast-update formulas:
\begin{eqnarray}
&R_{i,j} = G_{i,n}M_{n,j}, \\
&L_{i,j} = M_{i,n}G_{n,j}, \nonumber \\
&M^{(k)}=D^{-1 (k)}, \nonumber \\
&\Delta=M^{-1 (k+1)}-M^{-1 (k)}. \nonumber
\end{eqnarray}
Hereafter summation over repeated indices is implied and $(k)$
denotes size of the matrix. In the last formulae matrix $M^{(k)}$
is extended to be a $(k+1)\times (k+1)$ matrix with
$M_{k+1,k+1}=1$ and $M_{k+1,i}=0$, $M_{i,k+1}=0$ (it does not
change the ratio of determinants). Thus
\begin{eqnarray}
&M^{(k+1)}=M^{(k)}[1+\Delta M^{(k)}]^{-1},\\
&\frac{\det D^{(k+1)}}{\det D^{(k)}}=\frac{\det M^{(k)}}{\det M^{(
k+1)}}=\det [1+\Delta M^{(k)}]=\lambda. \nonumber
\end{eqnarray}
Using the standard $2\times 2$ super-matrix manipulations one can
obtain the following expression for $[1+\Delta M^{(k)}]^{-1}$
matrix:
\begin{widetext}
\begin{equation}
\lbrack 1+\Delta M^{(k)}]^{-1}=\left(
\begin{array}{ccccc}
1+G_{1,k+1}\lambda^{-1} R_{k+1,1} & G_{1,k+1}\lambda^{-1}
R_{k+1,2} & ... &
G_{1,k+1}\lambda^{-1} R_{k+1,k} & -G_{1,k+1}\lambda^{-1}  \\
G_{2,k+1}\lambda^{-1} R_{k+1,1} & 1+G_{2,k+1}\lambda^{-1}
R_{k+1,2} & ... &
G_{2,k+1}\lambda^{-1} R_{k+1,k} & -G_{2,k+1}\lambda^{-1}  \\
... & ... & ... & ... & ... \\
G_{k,k+1}\lambda^{-1} R_{k+1,1} & G_{k,k+1}\lambda^{-1} R_{k+1,2}
& ... &
1+G_{k,k+1}\lambda^{-1} R_{k+1,k} & -G_{k,k+1}\lambda^{-1}  \\
-\lambda^{-1} R_{k+1,1} & -\lambda^{-1} R_{k+1,2} & ... &
-\lambda^{-1} R_{k+1,k} & \lambda^{-1}
\end{array}
\right)
\end{equation}
\end{widetext}
Then it's easy to the obtain fast-update formulas for the step
$k+1$. Matrix $M^{(k+1)}$ can be obtained from $M^{(k)}$. Finally
the expressions for the matrix $M^{(k+1)}$ and for the ratio of
determinants have the following form:
\begin{widetext}
\begin{eqnarray}
&M^{(k+1)} =\left(
\begin{array}{cccc}
... & ...
& ... & -L_{1,k+1}\lambda^{-1}  \\
... & M'_{i,j} & ... & ... \\
... & ... & ... & -L_{k,k+1}\lambda^{-1}  \\
-\lambda^{-1} R_{k+1,1} & ... & -\lambda^{-1} R_{k+1,k} &
\lambda^{-1}
\end{array}
\right), \\
&M'_{i,j}=M^{(k)}_{i,j}+L_{i,k+1}\lambda^{-1} R_{k+1,j}, \nonumber \\
&\frac{\det D^{(k+1)}}{\det D^{(k)}}
=G_{k+1,k+1}-G_{k+1,i}M_{i,j}^{(k)}G_{j,k+1} = \lambda =
\frac{1}{M_{k+1,k+1}^{(k+1)}}, \nonumber
\end{eqnarray}
\end{widetext}
where $i,j=1,...,k$. For the step $k-1$ (removal of the column and
row $n$) the fast update formulas for matrix $M^{(k-1)}$ and the
ratio of determinants are as follows:
\begin{eqnarray}
&M_{i,j}^{(k-1)}=M_{i,j}^{(k)}-\frac{M_{i,n}^{(k)}M_{n,j}^{(k)}}{M_{n,n}^{(k)}%
}, \\
&\frac{\det D^{(k-1)}}{\det D^{(k)}}=\frac{\det M^{(k)}}{\det
M^{(k-1)}}=M_{n,n}^{(k)}. \nonumber
\end{eqnarray}

One can also obtain fast-update formulas in the same manner for
steps $k \pm 2$. Let's introduce a $2\times 2$ matrix $\lambda$:
\begin{equation}
\lambda_{q,q'}=G_{q,q'}-G_{q,i}M_{i,j}G_{j,q'}.
\end{equation}
where $q,q'=k+1,k+2$. Then the fast-update formulas for a step
$k+2$ look like
\begin{widetext}
\begin{eqnarray}
&M^{(k+2)} =\left(
\begin{array}{ccccc}
... & ...
& ... & -L_{1,q}\lambda_{q,k+1}^{-1} & -L_{1,q}\lambda_{q,k+2}^{-1} \\
... & M'_{i,j} & ... & ... & ... \\
... & ... & ... & -L_{k,q}\lambda_{q,k+1}^{-1} & -L_{k,q}\lambda_{q,k+2}^{-1} \\
-\lambda_{k+1,q'}^{-1} R_{q',1} & ... & -\lambda_{k+1,q'}^{-1}
R_{q',k} & \lambda_{k+1,k+1}^{-1} & \lambda_{k+1,k+2}^{-1} \\
-\lambda_{k+2,q'}^{-1} R_{q',1} & ... & -\lambda_{k+2,q'}^{-1}
R_{q',k} & \lambda_{k+2,k+1}^{-1} & \lambda_{k+2,k+2}^{-1}
\end{array}
\right), \\
& M'_{i,j}=M^{(k)}_{i,j}+L_{i,q}\lambda_{q,q'}^{-1} R_{q',j}, \nonumber \\
&\frac{\det D^{(k+2)}}{\det D^{(k)}}=\det \lambda, \nonumber
\end{eqnarray}
\end{widetext}
where $i,j=1,...,k$. For the step $k-2$ (removal of two columns
and two rows $n+1,n+2$) matrix $\lambda$ has the following form:
\begin{equation}
\lambda_{q,q'}=M_{q,q'}.
\end{equation}
where $q,q'=n+1,n+2$. Then the fast update formulas for the matrix
$M^{(k-2)}$ and the ratio of determinants are as follows:
\begin{eqnarray}
&M_{i,j}^{(k-2)}=M_{i,j}^{(k)}-M_{i,q}^{(k)}\lambda_{q,q'}^{-1}M_{q',j}^{(k)}, \\
&\frac{\det D^{(k-2)}}{\det D^{(k)}}=\frac{\det M^{(k)}}{\det M^{(
k-2)}}=\det [\lambda].\nonumber
\end{eqnarray}

Using the fast update formula for ratio of determinants, the Green
function can be obtained both in imaginary time and at Matsubara
frequencies:
\begin{eqnarray}
&g_{\tau'}^{\tau}=g_{0\tau'}^{\tau}-\sum\limits_{i,j} g_{0\tau_i}^{\tau} M_{i,j} g_{0\tau'}^{\tau_j},  \\
&g(i\omega)=g_0(i\omega)-g_0(i\omega)[\frac{1}{\beta}\sum\limits_{i,j}
M_{i,j} e^{i \omega (\tau_i - \tau_j)}]g_0(i\omega).\nonumber
\end{eqnarray}
Here $g_0(i\omega)$ is a bare Green function.

Higher correlators can be obtained from Wick's theorem, just as in
the auxiliary-field quantum Monte Carlo \cite{Hirsch} scheme. Also
note that it's convenient to keep in memory only the inverse
matrices $M$ instead of direct $D$ in simulations.

\subsection{The sign problem}

A proper choice of $\alpha$ can completely suppress the sign
problem in certain cases. To be concrete, let us consider a
Hubbard model. In this model the interaction is local in time and
space, and only electrons with opposite spins interact. Therefore
it is reasonable to take $\alpha_{t' i' \uparrow}^{t i \uparrow}
=\delta(\tau-\tau') \delta(i-i') \alpha_{\uparrow}$, similar for
$\alpha_\downarrow$, and
$\alpha^\downarrow_\uparrow=\alpha_\downarrow^\uparrow=0$. The
perturbation $W$ becomes
\begin{equation}\label{Hubbard}
W_{Hubbard}=U \int
(n_\uparrow(\tau)-\alpha_\uparrow)(n_\downarrow(\tau)-\alpha_\downarrow)
dt
\end{equation}
Here the Hubbard $U$ and the occupation number operator $n=c^\dag
c$ are introduced. The Gaussian part of the Hubbard action is
spin-independent and does not rotate spins. This means that only
$g_\downarrow^\downarrow, g^\uparrow_\uparrow$ do not vanish, and
the determinant in (\ref{det}) is factorized
\begin{equation}\label{updown}
D^{r_1 r_2 ... r_{2k}}_{r_1' r_2' ... r_{2k}'}=D^{r_1 r_3 ...
r_{2k-1}}_{r_1' r_3' ... r_{2k-1}'} D^{r_2 r_4 ... r_{2k}}_{r_2'
r_4' ... r_{2k}'} \equiv D_\uparrow D_\downarrow
\end{equation}
For the case of attraction $U<0$ one should choose
\begin{equation}\label{attract}
    \alpha_\uparrow=\alpha_\downarrow=\alpha,
\end{equation}
where $\alpha$ is a real number. For this choice
$g_\downarrow^\downarrow=g^\uparrow_\uparrow$, and therefore
$D_\uparrow=D_\downarrow$. $\Omega$ is always positive in this
case, as follows from formula (\ref{Z}).

This choice of $\alpha$ is useless for a system with repulsion,
however. Compared to the case of attraction, another sign of $w$
at $\alpha_\uparrow=\alpha_\downarrow$ results in alternating
signs of $\Omega_k$ with odd and even $k$ \cite{Alt}. Another
condition for $\alpha$ is required. The particle-hole symmetry can
be exploited for the Hubbard model at half-filling. In this case,
the transformation $c^\dag_\downarrow \to \tilde{c}_\downarrow$
converts the Hamiltonian with repulsion to the same but with
attraction. Therefore the series (\ref{Z}) in powers of $W=U \int
(n_\uparrow(\tau)-\alpha)(\tilde{n}_\downarrow(\tau)-\alpha) dt$
does not contain negative numbers, in accordance to the previous
paragraph. The value of the trace in (\ref{Z}) is independent of a
particular representation. In the original (untransformed) basis
the above $W$ reads as $U \int (n_\uparrow(\tau)-\alpha)
(n_\uparrow(\tau)-1+\alpha) dt$. We conclude that
\begin{equation}\label{repuls}
\alpha_\uparrow=1-\alpha_\downarrow=\alpha
\end{equation}
eliminates the sign problem for repulsive systems with a
particle-hole symmetry. Of course, the average sign for a system
with repulsion is not equal to unity in a general case.

It is useful to analyze a toy single-atom Hubbard model to get a
feeling for the behavior of the series (\ref{Z}). The two parts of
the action are
\begin{eqnarray}
&S_0=\int  (-\mu+U \alpha_\downarrow)
n_\uparrow(\tau)+(-\mu+U \alpha_\uparrow) n_\downarrow(\tau)) d\tau;\\
\nonumber &W=U \int (n_\downarrow(\tau)-\alpha_\downarrow)
(n_\uparrow(\tau)-\alpha_\uparrow) d\tau.
\end{eqnarray}
Here $\mu$ is a chemical potential. For a half-filled system
$\mu=U/2$. For this model, it is easy to calculate terms of the
series for $Z$ explicitly. We obtain
\begin{eqnarray}
&\Omega_k= \frac{(-U \alpha_\uparrow
\alpha_\downarrow)^k}{k!}\left(1+e^{\beta (\mu-U
\alpha_\downarrow)}(1-\alpha_\uparrow^{-1})^k\right)\times
\nonumber \\
&\times \left(1+e^{\beta (\mu-U
\alpha_\uparrow)}(1-\alpha_\downarrow^{-1})^k\right).
\end{eqnarray}
Consider the case of repulsion ($U>0$). Let us use the condition
(\ref{repuls}) for an arbitrary filling factor. The later
expression can be presented in the form
\begin{eqnarray}\label{imp}
&\Omega_k=e^{\beta (\mu-U \alpha)} \frac{(U \alpha^2)^k}{k!}
\left(1+e^{\beta (\mu-U+U \alpha)}(1-\alpha^{-1})^k\right)\times
\nonumber \\
        &\times \left(1+e^{\beta (-\mu+U
        \alpha)}(1-\alpha^{-1})^k\right).
\end{eqnarray}
For $\mu=U/2$ the value of $\Omega_k$ is positive for any
$\alpha$. For a general filling factor, the situation depends on
the value of $\alpha$. For $0<\alpha<1$ negative numbers can occur
at certain $k$. Outside this interval all terms are positive, and
there is no sign problem for the single-atom system under
consideration.

Since  the sign problem exists already for the impurity problem
for $0<\alpha<1$, such a choice is also not suitable  for the
$N$-atom repulsive Hubbard system. On the other hand, minimization
of $\bar{W}$ requires $\alpha$ to be as close to this interval as
possible. Therefore it is reasonable to take $\alpha=1$ or
slightly above. This is the same as zero or a small negative
value, since $\alpha_\uparrow=1-\alpha_\downarrow$. We use the
similar choice of $\alpha$'s for more complicated multi-orbital
models and always obtain a reasonable average sign.

Finally, one may prefer to have a perturbation that is symmetrical
in spin projections. Formula (\ref{attract}) for the attractive
interaction is already symmetrical. For the case of repulsion we
propose to use a symmetrized form
\begin{equation}\label{Wrepuls}
    \frac{U}{2}(n_{\uparrow} + \alpha) (n_{\downarrow} -1
    -\alpha)+\frac{U}{2}(n_{\uparrow} -1- \alpha) (n_{\downarrow}
    +\alpha)
\end{equation}
with some small positive $\alpha$.

There is another argumentation why the presence of $\alpha$'s in
Eq. (\ref{Wrepuls}) is very important. Indeed, proper choice of
$\alpha$ make the average of Eq. (\ref{Wrepuls}) negative. We can
call such an interaction "virtually attractive in average". It
makes possible to obtain the $k$-series with the all-positive
integrals in the expansion, whereas the same series without
$\alpha$'s is useless due to the alternative signs of integrals.
We believe that the similar reasoning is valid for the non-local
interaction. Note however that the proper choice of the $\alpha$'s
depends on the particular system under calculation. For now, we
cannot offer a general recipe. In a certain situation, the
expressions under the integrals are not always positive, and the
exponential falloff occurs for the large systems or small
temperature. The practical calculations of the average sign and
comparison with the discrete-time QMC scheme are presented in the
next Section.

\section{Applications of CT-QMC method}

We test present algorithm for several well known models in this
Section. These examples show some of the advantages of the CT-QMC
method.

In all examples presented below we calculate a Green function at
Matsubara frequencies $G(i\omega_n)$. Total number of Matsubara
frequencies is varied from $10$ to $20$. The typical number of QMC
trials is $10^6 \div 10^7$. Normally, the error bar of the CT-QMC
data for $G(i\omega_n)$ is less than $3\cdot 10^{-3}$ for the
lowest Matsubara frequency and becomes smaller as frequency
increases. Obviously, values of these typical parameters depend on
concrete system.

\subsection{Hubbard clusters}

To test the scheme, we start from a single isolated Hubbard atom
and a $2\times 2$ Hubbard cluster. Results are compared with the
known exact solution (see e.g. Ref. \onlinecite{DMFT}).

\begin{figure}
\includegraphics[width=\columnwidth]{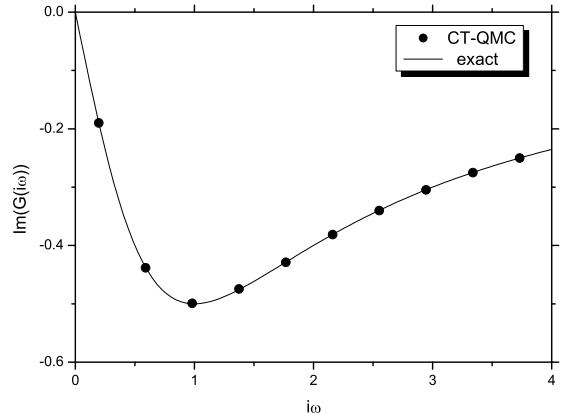}
\caption{\label{singleatom} Imaginary part of the Green function
at Matsubara frequencies for a single atom with Hubbard repulsion
$U$. Symbols are CT-QMC data, line is an exact solution
\cite{DMFT}. Parameters: $U=2, \beta=16, \mu=U/2$. Error bar is
less than symbol size.}
\end{figure}

The  solution for the atomic limit reads as follows:
\begin{eqnarray}
&G(i\omega)=\frac{1-n}{i\omega + \mu}+ \frac{n}{i\omega + \mu - U}, \\
&n=(e^{\beta \mu}+e^{\beta(2\mu - U)}) / (1+2e^{\beta
\mu}+e^{\beta(2\mu - U)}). \nonumber
\end{eqnarray}
Results for $U=2, \beta=16, \mu=U/2$ are presented in
Figure~\ref{singleatom}. Thus CT-QMC data are in an excellent
agreement with the analytical solution.

\begin{figure}
\includegraphics[width=\columnwidth]{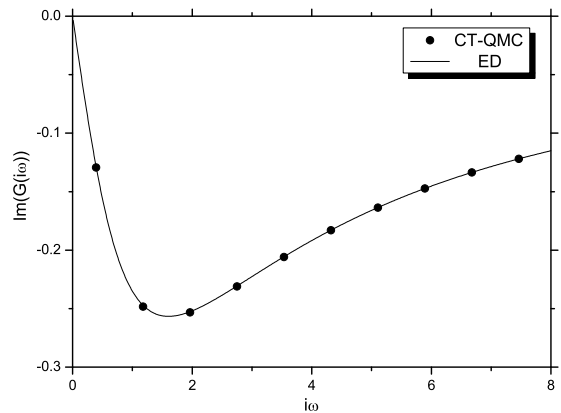}
\caption{\label{22Hubbardhf} $2\times 2$ Hubbard lattice at
half-filling. Imaginary part of the Green function at Matsubara
frequencies: symbols are CT-QMC data, line is
exact-diagonalization data. Parameters: $U=4, t=1, \beta=8,
\mu=U/2$. Error bar is less than symbol size.}
\end{figure}

Further we apply CT-QMC algorithm to the $2\times 2$ Hubbard
lattice to compare with the  auxiliary-field quantum Monte Carlo
scheme \cite{Hirsch}. We start with the half-filled case
($\mu=U/2$, four electrons in the system). It can be shown that
for the particular case of half-filling one can choose
$\alpha_\uparrow=\alpha_\downarrow=0.5$ due to the particle-hole
symmetry. Expression (\ref{Kaver}) for this case becomes
$<k>=\beta N (0.5-<n_{\uparrow} n_{\downarrow}>)$ with $N=4$. It
can be verified that this choice  delivers the minimal possible
$<k>$. Series (\ref{Z}) contains only the terms with an even $k$
in this case, so it's appropriate to use steps $\pm 2$. Results
for $U=4, t=1, \beta=8$ in comparison with the
exact-diagonalization data are shown in Figure~\ref{22Hubbardhf}.

\begin{figure}
\includegraphics[width=\columnwidth]{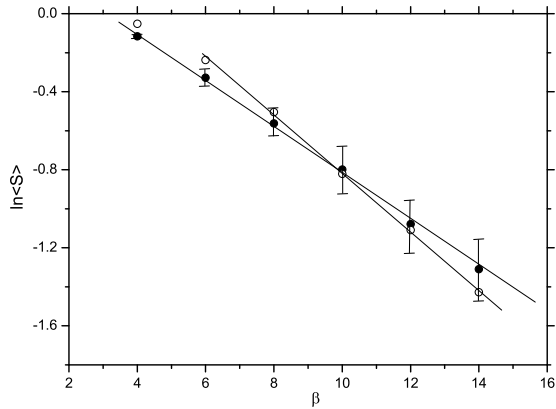}
\caption{\label{22Hubbardsign} $2\times 2$ Hubbard lattice away
from half-filling: three electrons in the system. Average sign as
a function of $\beta$: CT-QMC (filled symbols) and auxiliary-field
quantum Monte Carlo \cite{Hirsch} (opened symbols) algorithms
results. Lines are guides to the eye. Parameters: $U=4, t=1$.}
\end{figure}

\begin{figure}
\includegraphics[width=\columnwidth]{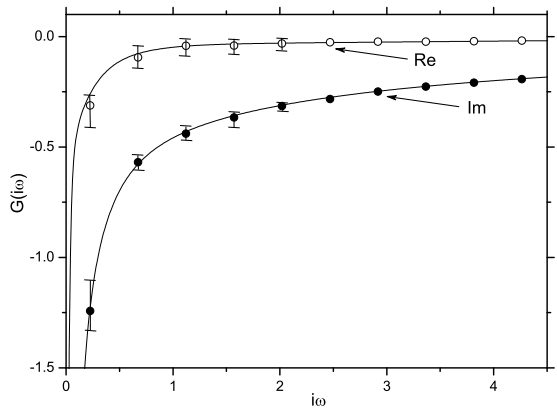}
\caption{\label{22Hubbardb14} Real and imaginary parts of the
Green function for $2\times 2$ Hubbard lattice away from
half-filling: three electrons in the system. Parameters: $U=4,
t=1, \beta=14$. Symbols are CT-QMC data, lines are
exact-diagonalization results. Error bar for $i\omega>2$ is less
than symbol size.}
\end{figure}

Cases of a single atom and a half-filled cluster do not suffer a
sign problem. One can discuss a sign problem considering $2\times
2$ Hubbard lattice away from half-filling. For this case a choice
(\ref{Wrepuls}) for $\alpha$'s was used. We concentrate on the
worst sign-problem case when there are three electrons in the
system \cite{sign2x2}. The average sign is presented in
Figure~\ref{22Hubbardsign} as a function of inverse temperature
$\beta$. We would like to stress that the CT-QMC algorithm agrees
with the auxiliary-field quantum Monte Carlo \cite{Hirsch} scheme
(Figure~\ref{22Hubbardsign}). Even for a relatively small average
sign, numerical data remain to be in a good agreement with the
exact-diagonalization, as Figure~\ref{22Hubbardb14} shows.

\subsection{Metal-insulator transition on the Bethe lattice}

One of the advantages of the CT-QMC algorithm is a possibility to
perform simulations at lower temperatures with higher accuracy
than the auxiliary-field quantum Monte Carlo \cite{Hirsch} method.
Here we present results for the metal-insulator phase transition
in Hubbard model on Bethe lattice \cite{DMFT}. The effective
one-site problem based on the dynamical mean-field theory
\cite{DMFT} is solved by CT-QMC method.

The standard self-consistent loop of DMFT equations is as follows
\cite{DMFT}. One starts with some initial guess for the Green
function ${\cal G}_0$ which is used to obtain the local Green
function ${\cal G}$ from the effective action as \cite{DMFT}
\begin{equation} \label{dmft1}
{\cal G}(\tau,\tau')=<T c^\dag_{\tau'} c^{\tau}>_{S_{eff}({\cal
G}_{0})}.
\end{equation}
A new guess for the Green function ${\cal G}_{0}$ is obtained from
the equation for Bethe lattice ($t=1/2$) \cite{DMFT}:
\begin{equation} \label{dmft2}
{\cal G}_{0}^{-1}(i\omega)=i\omega + \mu - t^2 {\cal G}(i\omega).
\end{equation}
Formulas (\ref{dmft1},\ref{dmft2}) form a self-consistent loop of
DMFT equations. The Green function which corresponds to the
semi-circular density of states with band-width 2 is usually used
for the Bethe lattice:
\begin{equation} \label{dmft3}
{\cal G}_0(i\omega)=\frac{2}{i(\omega + \sqrt{\omega ^2+1})+
2\mu}.
\end{equation}
The self-energy $\Sigma(i\omega)$ can be obtained from the
following formula after the iteration procedure for the DMFT
equations (\ref{dmft1},\ref{dmft2}) has converged:
\begin{equation} \label{dmft4}
\Sigma(i\omega)={\cal G}_{0}^{-1}(i\omega)-{\cal G}^{-1}(i\omega).
\end{equation}

\begin{figure}
\includegraphics[width=\columnwidth]{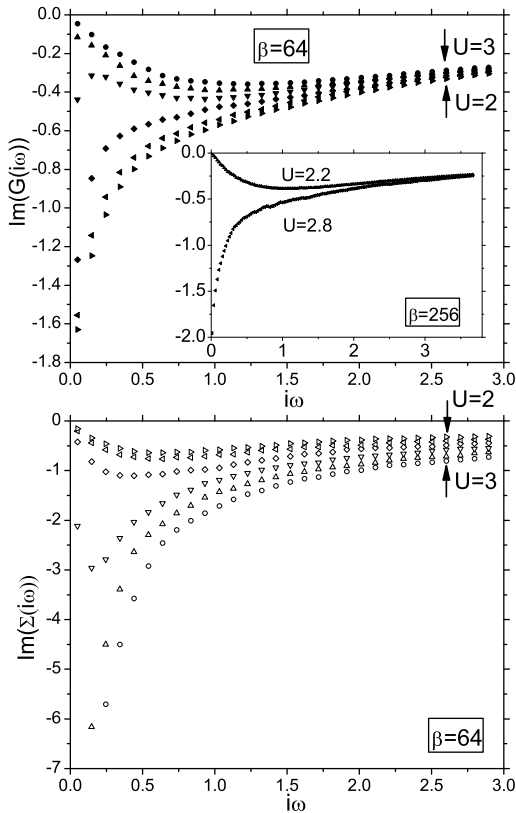}
\caption{\label{MIT} Imaginary part of the Green function (a) and
self-energy (b) at Matsubara frequencies for Hubbard model on
Bethe lattice at half-filling obtained from the solution of
self-consistent DMFT equations (\ref{dmft1},\ref{dmft2}) by CT-QMC
method. Parameters: $\beta=64, U=2 \div 3, \Delta U=0.2$. All data
obtained with the initial guess for the Green function in the form
(\ref{dmft3}) which corresponds to the metallic phase. Coexistence
of metallic and insulating phases can be found, for example, at
point $U=2.4$. Inset shows data for the imaginary part of the
Green function for $\beta=256$, $U=2.2$ and $U=2.8$.}
\end{figure}

Results for the metal-insulator phase transition in Hubbard model
on Bethe lattice at half-filling for $\beta=64$ are presented in
Figure~\ref{MIT}. Local Green functions and corresponding
self-energies are shown for values of Coulomb interaction $U$ from
the value $U=2$ to the value $U=3$ with the step $\Delta U=0.2$.
The results show a phase transition from the metallic state
(smaller values of $U$) to the insulating state (larger values of
$U$) with a coexistence region in between. The data obtained agree
well with previous studies of the transition where the standard
auxiliary-field quantum Monte Carlo \cite{Hirsch} algorithm was
used as a solver for DMFT equations (\ref{dmft1},\ref{dmft2})
\cite{DMFT}. Note, CT-QMC scheme gives better accuracy than
auxiliary-field quantum Monte Carlo \cite{Hirsch} algorithm since
one obtains the local Green function at Matsubara frequencies
directly in QMC. It allows one to perform simulations at lower
temperatures. For instance, we tested the CT-QMC algorithm even at
$\beta=256$ and obtained quite reasonable results for the
metal-insulator phase transition on Bethe lattice (see inset for
Figure~\ref{MIT} as well).

\subsection{Multi-band model with a rotationally-invariant retarded exchange}

Another advantage of the CT-QMC algorithm is that it allows one to
consider multi-band problems with interactions in the most general
form:
\begin{equation} \label{genint}
\hat{U}=\frac {1}{2} \sum_{ijkl;\sigma
\sigma'}U_{ijkl}c^\dag_{i\sigma}c^\dag_{j\sigma'}c^{l\sigma'}c^{k\sigma}.
\end{equation}

\begin{figure}
\includegraphics[width=\columnwidth]{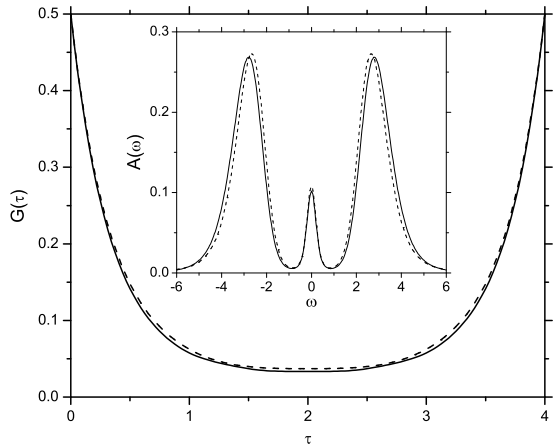}
\caption{\label{rot_inv} Imaginary-time Green Function for the
rotationally-invariant two-band model. Solid and dot lines
correspond to the static and to the nonlocal in time spin-flip,
respectively. The inset shows DOS estimated from the Green
function.}
\end{figure}

\begin{figure}
\includegraphics[width=\columnwidth]{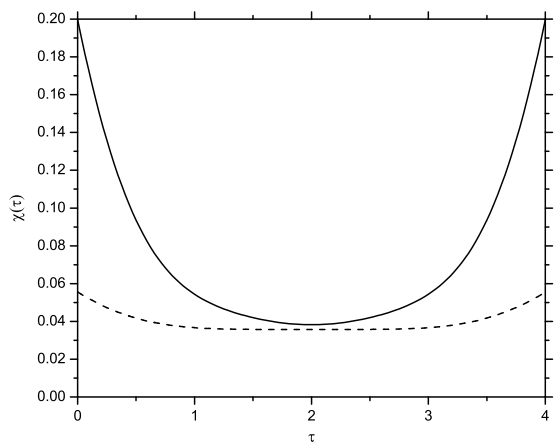}
\caption{\label{rot_inv_chi} Imaginary-time dependence of the
four-point quantity $\chi(\tau-\tau^{\prime})= <c^{\dagger}_{0
\uparrow \tau} c^{0 \downarrow \tau} c^{\dagger}_{1 \downarrow
\tau^{\prime}} c^{1 \uparrow \tau^{\prime}}>$ for the
rotationally-invariant two-band model. Solid and dot lines
correspond to the static and to the nonlocal in time spin-flip,
respectively.}
\end{figure}

We apply the proposed CT-QMC for the important problem of the
super-symmetric two band impurity model at half-filling
\cite{Rozenberg, Dworin}. To our knowledge, this is the first
successful attempt to take the off-diagonal exchange terms of this
model into account. These terms are important for the realistic
study of the multi-band Kondo problem because they are responsible
for the local moment formation \cite{Dworin}. The interaction in
this model has the following form
\begin{equation}
\frac{U}{2} (\hat{N}(\tau)-2) (\hat{N}(\tau)-2) - \frac{J}{2}
({\bf S}(\tau) \cdot {\bf S}(\tau)+ {\bf L}(\tau) \cdot {\bf
L}(\tau)),
\end{equation}
where $\hat{N}$ is the operator of total number, $S$ and $L$ are
total spin and orbital-momentum operators, respectively. The
interaction is spin- and orbital- rotationally invariant. The
Gaussian part of the action represents the diagonal semicircular
density of states \cite{DMFT} with unitary half-band width
(\ref{dmft3}). We used parameters $U=4, J=1$ at $\beta=4$.
Figures~\ref{rot_inv} and \ref{rot_inv_chi} present the results
for the local Green function $G_{i s}^{i s} (\tau)$ and the
four-point correlator $\chi(\tau-\tau^{\prime})= <c^{\dagger}_{0
\uparrow \tau} c^{0 \downarrow \tau} c^{\dagger}_{1 \downarrow
\tau^{\prime}} c^{1 \uparrow \tau^{\prime}}>$. The later quantity
characterizes the spin-spin correlations and would vanish if the
exchange were absent.

A modification of this model was also studied where spin-flip
operators were replaced with the terms fully non-local in time.
For example, operator $c^{\dagger}_{0 \uparrow \tau} c^{0
\downarrow \tau} c^{\dagger}_{1 \downarrow \tau} c^{1 \uparrow
\tau}$  was replaced with $\beta^{-1}\int d\tau^{\prime}
c^{\dagger}_{0 \uparrow \tau} c^{0 \downarrow \tau} c^{\dagger}_{1
\downarrow \tau^{\prime}} c^{1 \uparrow \tau^{\prime}}$. As it is
pointed in the introduction, the retardation effects in the
interaction always appear if certain non-Gaussian degrees of
freedom are integrated out. Therefore it is of importance to
demonstrate that CT-QMC scheme is able to handle the retarded
interaction.

The Green function in the time domain was obtained by a numerical
Fourier-transform from the CT-QMC data for $G(i \omega_n)$. For
high harmonics the following asymptotic form was used: $-{\rm Im}
(i w+ \epsilon)^{-1}$ with $\epsilon \approx 2.9$. The obtained
dependencies are presented in Figure~\ref{rot_inv}. Results for
the local and non-local in time spin-flip interactions are shown
with solid and dot lines, respectively. It is interesting to note
that the Green function is rather insensitive to the details of
spin-flip retardation. The maximum-entropy guess for DOS is
presented in the inset to Figure~\ref{rot_inv}. Both Green
functions are very similar and correspond to qualitatively the
same density of states (DOS).

To demonstrate the effects due to retardation we calculated the
four-point quantity $\chi(\tau)$. These data are obtained
similarly, the difference is that $\chi(i \omega)$ is defined at
Bose Matsubara frequencies and obeys a $1/\omega^2$ decay. It
turns out that a switch to the non-local in time exchange modifies
$\chi(\tau)$ dramatically. The local in time exchange results in a
pronounced peak of $\chi(\tau)$ at $\tau \approx 0$, whereas the
non-local spin-flip results in almost time-independent spin-spin
correlations (Figure~\ref{rot_inv_chi}).

\section{Concluding remarks}

In conclusion, we have developed a fermionic continuous time
quantum Monte Carlo method for general non-local in space-time
interactions. It's successfully tested for a number of models.

We demonstrated that for Hubbard-type models the computational
time for a single trial step scales similarly to that for the
schemes based on a Stratonovich transformation. An important
difference occurs however for the non-local interactions.
Consider, for example, a system with a large Hubbard $U$ and much
smaller but still important Coulomb interatomic interaction. One
needs to introduce $N^2$ auxiliary fields per time slice instead
of $N$ to take the long-range forces into account. On the other
hand, the complexity of the present algorithm remains almost the
same as for the local interactions, because $\overline{|W|}$ does
not change much. This should be useful for the realistic cluster
DMFT calculations and for the applications to quantum chemistry
\cite{qchem}. It is also possible to study the interactions
retarded in time, particularly the super-exchange and the effects
related to dissipation. This was demonstrated for an important
case of the fully rotationally invariant multi-band model and its
extension with non-local in time spin-flip terms.

For the case of the Hubbard model the sign problem was found to be
similar to what occurs for the auxiliary-field quantum Monte Carlo
\cite{Hirsch} scheme. Nevertheless a general time-dependent form
of the action (Eq.(\ref{S0})) opens, in principal, the possibility
for a two-stage renormalization treatment. Suppose we know a
certain renormalization of action, based on the local
DMFT-solution as a starting point. Since DMFT is already a very
good approximation, we can expect the thus renormalized
interaction to be smaller than the initial one, although it is
perhaps nonlocal in time. Then one could expect that the lattice
calculations with a renormalized interaction show a smaller sign
problem. Practical investigation of such constructed
renormalization is a subject of the future work.

We are grateful to A. Georges, M. Katsnelson and F. Assaad for
their very valuable comments. This research was supported in part
by the National Science Foundation under Grant No. PHY99-07949,
"Russian Scientific Schools" Grant 96-1596476, and FOM Grant
N0703M. Authors (AR, AL) would like to acknowledge a hospitality
of KITP at Santa Barbara University and (AR) University of
Nijmegen. The CT-QMC program described in this article are
available at http://www.ct-qmc.ru or via e-mail (AR, alex@shg.ru).

\end{document}